\documentclass[10pt,letterpaper]{article}
\usepackage[top=0.85in,left=2.75in,footskip=0.75in]{geometry}

\usepackage{amsmath,amssymb}

\usepackage{changepage}

\usepackage{textcomp,marvosym}

\usepackage{cite}

\usepackage{nameref,hyperref}

\usepackage[right]{lineno}

\usepackage[nopatch=eqnum]{microtype}
\DisableLigatures[f]{encoding = *, family = * }

\usepackage[table]{xcolor}

\usepackage{array}

\newcolumntype{+}{!{\vrule width 2pt}}

\newlength\savedwidth

\newcommand\thickhline{\noalign{\global\savedwidth\arrayrulewidth\global\arrayrulewidth 2pt}%
\hline
\noalign{\global\arrayrulewidth\savedwidth}}

\raggedright
\usepackage[aboveskip=1pt,labelfont=bf,labelsep=period,justification=raggedright,singlelinecheck=off]{caption}

\makeatletter
\renewcommand{\@biblabel}[1]{\quad#1.}
\makeatother

\usepackage{lastpage,fancyhdr,graphicx}
\usepackage{epstopdf}
\fancyheadoffset[L]{2.25in}
\fancyfootoffset[L]{2.25in}
\begin{document}
\vspace*{0.2in}

\begin{flushleft}
{\Large
\textbf\newline{Discounting the distant future:
What do historical bond prices imply about the long term discount rate?} 
}
\newline
\\
J. Doyne Farmer\textsuperscript{1,2},
John Geanakoplos\textsuperscript{2,3},
Matteo G. Richiardi\textsuperscript{4},
Miquel Montero\textsuperscript{5,6*},
Josep Perell\'o\textsuperscript{5,6*},
Jaume Masoliver\textsuperscript{5,6*}
\\
\bigskip
\textbf{1} Institute for New Economic Thinking at the Oxford Martin School, University of Oxford, Oxford OX1 3UQ, United Kingdom 
\\
\textbf{2} Santa Fe Institute, Santa Fe NM 87501, United States
\\
\textbf{3} Department of Economics, Yale University, New Haven CT 06511, United States
\\
\textbf{4} Institute for Social and Economic Research, University of Essex, Wivenhoe Park CO4 3SQ, United Kingdom 
\\
\textbf{5}
Departament de F\'isica de la Mat\`eria Condensada, Universitat de Barcelona
\\
\textbf{6}
Universitat de Barcelona Institute of Complex Systems Institute (UBICS), Barcelona E-08028, Spain
\bigskip

%
%





* miquel.montero@ub.edu, josep.perello@ub.edu, jaume.masoliver@ub.edu

\end{flushleft}
\section*{Abstract}
We present a thorough empirical study on real interest rates by also including risk aversion through the introduction of the market price of risk. With the view of complex systems science and its multidisciplinary approach, we use the theory of bond pricing to study the long term discount rate. Century-long historical records of 3 month bonds, 10 year bonds, and inflation allow us to estimate real interest rates for the UK and the US. Real interest rates are negative about a third of the time and the real yield curves are inverted more than a third of the time, sometimes by substantial amounts.  This rules out most of the standard bond pricing models, which are designed for nominal rates that are assumed to be positive. We therefore use the Ornstein-Uhlenbeck model which allows negative rates and gives a good match to inversions of the yield curve. We derive the discount function using the method of Fourier transforms and fit it to the historical data. The estimated long term discount rate is $1.7$ \% for the UK and $2.2$ \% for the US. The value of $1.4$ \% used by Stern is less than a standard deviation from our estimated long run return rate for the UK, and less than two standard deviations of the estimated value for the US. All of this once more reinforces the support for immediate and substantial spending to combat climate change.


\section{Introduction}

The evaluation of the economic impact of global warming requires a balance of future costs against present costs \cite{DasGupta2004}. The most basic assumption considers an exponential with a constant discount rate.  The choice of discount rate is part on current economic debates  on the urgency of the response to global warming (see \cite{ArrowReviewScience,Stern,Nordhaus,Nordhaus2007}). Most normative approaches attempts to derive the discount from axiomatic principles of justice, or from utility theory and assumptions about growth \cite{Stern1,Stern2}. Discounting includes impatience, economic growth, and declining marginal utility and it can embedded in the Ramsey formula \cite{Drupp2018,Ramsey,ArrowReview}). This approach makes however difficult to make quantitative estimates because factors involved are difficult to be measured empirically.

Here we take a positive approach assuming that  discounting is equivalent to bond pricing. However, time to maturity of bonds is not generally higher than 30 years, but climate change time horizons need the discount 100 years or even more into the future. Thus, inferring long maturity bonds prices from empirical data on short maturity bonds is fundamental and long historical time series for having a reliable statistical inference are needed. This data is most often not available and then it will be required to consider a reasonable model for real interest rates at different maturities.

There are two effects influencing long term rates that must be considered. The first of these is risk aversion. Longer term bonds bear greater risk than shorter ones, and higher risk should imply higher interest rates. The second effect is more subtle, and is due to the fact that interest rates are uncertain and highly persistent. In the environmental context, this effect was originally pointed out by Weitzman \cite{Weitzman98} and Gollier et al. \cite{Gollier} (see also \cite{Freeman_Groom(2016)}). However, the effect was also pointed out in a general context in \cite{Dybvig} and it has been known much longer in the context of bond pricing \cite{Vasicek}. The long run rate is thus dominated by the lowest value, since asymptotically all the other discount factors will be negligible in comparison. To see this concretely, consider two possible rates, $r_1$ and $r_2$, with $r_1 > r_2$, and assume they have equal probability.  Then the average discount factor at time $t$ in the future is $D(t) = (e^{-r_1 t} + e^{-r_2 t})/2$.   Note that since the sum of two exponentials is not an exponential, the discounting function is no longer an exponential.  But for $t$ sufficiently large $D(t) \approx e^{-r_2 t}/2$, i.e. the discount function becomes approximately exponential with the lower interest rate.  This illustrates that when interest rates are uncertain and persistent, lower interest rates tend to dominate.

We are here building on earlier empirical work.  Stochastic interest rates provided by Litterman et al. \cite{Litterman}, Newell and Pizer \cite{NewellPizer}, and Groom et al. \cite{Groom} are more realistic processes than those provided by Weitzman and Gollier. Groom et al. \cite{Groom} also noted that the drop in rates depends on uncertainty and persistence. All these authors calibrated the models suggested against long historical time series of 10 year bonds. We hare add to these previous works in several ways. We study century-long records of the prices of 3 month bonds, 10 year bonds, and inflation, for both the US and the UK.  Real interest rates are found to become negative more than one third of the time, often by substantial amounts. To our knowledge only few authors have addressed this issue  (e.g. Freeman et al \cite{Freeman_2015}) while the usual approach has been ignoring this fact, and instead forced historical interest rates to be positive in order to be consistent with their models. 

We instead take the view that negative real interest rates are a fact that cannot be ignored.  This leads us to choose the Ornstein-Uhlenbeck (OU) model, which is compatible with negative real interest rates. The model provides real interest rates that can be negative but they are constantly pulled back to certain rate. The formula for the yield curve for the OU model has been derived in the finance literature using several approaches (see, for instance, \cite{Duffie,Brigo,Piazzesi}; and the review in \cite{Mamon2004}). We derive this by a somewhat simpler way using Fourier transform methods. The previous work in environmental discounting cited above \cite{Groom} was based on numerical simulations. In contrast, we take advantage of the existence of closed form solutions, which allows us to better estimate the long term discount rate. When the OU model is used in the finance literature, the nominal interest rate and the mean reversion parameter is taken to be so strong that nominal interest rate goes negative with probability close to zero. In a different approach, Davidson et al. \cite{Davidson} examined a square root Ornstein Uhlenbeck model which can never go negative, is pulled down toward 0 as it is an absorbing state. Davidson et al. \cite{Davidson} has solved the asymptotic (long) rate by using the Feynman-Kac functional, which is quite different from our approach. 

The prevalence of negative real rates makes many of the standard nominal interest rate models inappropriate vehicles for studying real rates, and the OU becomes an obvious choice. In addition to our treatment of negative rates, we also give an original contribution regarding properly taking risk aversion into account. The works described above \cite{NewellPizer,Groom} were calibrated against a single maturity bond (10 years). Risk aversion is a well-accepted notion in bond pricing, and one would expect models that neglect it to underestimate long term interest rates. We fix this by taking risk aversion into account and fit the resulting model to both 3 month and 10 year bond yields, which provide us with two points on the yield curve.

In summary, we herein present a complete empirical study on real interest rates of US and UK for which we have century-long records and that are considered ones of the most economically stable countries as far inflation and economic growth are concerned. Our focus is based on the interdisciplinary approach of complex systems developments. We have also presented elsewhere a similar study for other less stable countries but with minor statistical reliability due to scarcity of data and without considering risk aversion (i.e., the market price of risk) \cite{perello_etal_JSM}. Several different theoretical aspects of discounting have been recently discussed by some of us mainly regarding the effects on discounting of extreme market situations \cite{mmp_mathemat_2021} or the introduction of resettings in real interest rates \cite{mpm_JPA_2022} (see also the recent review \cite{mmpdg_entropy_2022}).

\section{The process of discounting in continuous time}
\label{discounting}

We now derive the form of the yield curve for the OU model.  In the continuous limit the discount function is
\begin{equation}
D(t)=\mathbb{E}\left[\exp\left(-\int_{t_0}^{t} r(t^{\prime})dt^{\prime}\right)\right],
\label{discount_def}
\end{equation}
where $t_0$ is the initial time and the expectation $\mathbb E[\cdot]$ is an average over all possible real rate trajectories $r(t')$ $(t_0\leq t'\leq t)$  up to time $t$. The mathematical expression is formally identical to the bond prices. The price $B(t_0|t_0+ t)$ of a zero-coupon bond issued at time $t_0$ with unit payoff and time maturity $t_0+t$ 
($t \geq 0$) reads
\begin{equation}
B(t_0 |t_0 +t)=\mathbb{E}\left[\exp\left(-\int_{t_0}^{t} n(t^{\prime})dt^{\prime}\right)\right],
\label{bond_1}
\end{equation}
which depends on the nominal rate $n(t)$.

\subsection{The general framework}

Following the same strategy as for bonds, we compute $D(t)$ via a stochastic process model for the time evolution of $r(t)$. Simplest and most common assumption is to consider $r(t)$ as a diffusion process. The stochastic differential equation is
\begin{equation}
dr(t)=f(r)dt+ g(r) dW(t),
\label{sde_1}
\end{equation}
where $f(r)$ gives reason of the drift while $g(r)>0$ quantifies the noise intensity $dW(t)=\xi(t)dt$ (a Wiener process) characterized with
\begin{equation}
\mathbb{E}\left[\xi(t)\right]=0, \qquad \mathbb{E}\left[\xi(t)\xi(t')\right]=\delta(t-t'),
\label{xi_def}
\end{equation}
and where $\delta(\cdot)$ is Dirac's delta function.

The discount function is therefore given by
\[
D(t)=\mathbb E\left[  e^{-x(t)}\right]. 
\]
where the cumulative process $x(t)$ is defined as
\begin{equation}
x(t)=\int_{t_0}^{t}r(t^{\prime})dt^{\prime}.
\label{x}
\end{equation}

Therefore, the discount function can written in the following form:
\begin{equation}
D(t)=\int_{-\infty}^{\infty}dr\int_{-\infty}^{\infty}e^{-x}p(x,r,t|x_0,r_{0}, t_0)dx.
\label{D}
\end{equation}
The function $p(x,r,t|x_0,r_{0}, t_0)$ corresponds to the probability density function (PDF) that depends on $(x(t),r(t))$. The measure $p$ is often referred to as the {\it data generating measure}. We note that when there are uncertainties in the value of certain parameters which may appear in the estimation of the PDF from empirical data then the average in Eq (\ref{D}) has to be extended in order to include these additional uncertainties (see, e.g. \cite{Gollier2008}). 

Equations (\ref{sde_1}) and (\ref{x}) together refer to a bidimensional process defined in terms of a pair of stochastic differential equations
\begin{eqnarray}
dx(t)  &  = & r(t)dt, \label{dx}\\
dr(t)  &  = & f ( r) +g( r)dW(t), \label{dr}.
\end{eqnarray}
These stochastic different equations imply that the joint density follows the Fokker-Planck equation (FPE)
\begin{equation}
\frac{\partial p}{\partial t}=-r\frac{\partial p}{\partial x}-\frac{\partial}{\partial r}[f( r)p]+
\frac{1}{2}\frac{\partial^{2}}{\partial r^{2}}[g^2( r)p].
\label{fpe_1}
\end{equation}
The initial condition of this equation are fixed and are
\begin{equation}
p(x,r,t_0|x_0,r_{0},t_0)=\delta(x)\delta(r-r_{0}).
\label{initial_1}
\end{equation}

The process $r(t)$ is time homogeneous because $f(r)$ and $g(r)$ do not depend on time explicitly. The process is therefore time invariant and, without loss of generality, $t_0=0$.

\subsection{The Ornstein-Uhlenbeck process}

We now make explicit choices for the functions $f$ and $g$. For the reasons given in the introduction, we focus on the OU process, which is a standard model from finance that allows negative rates.  In finance the OU process was originally proposed by Vasicek \cite{Vasicek} and is sometimes referred to as the Vasicek model. It is a diffusion process with linear drift and constant noise intensity
\begin{equation}
f(r)=-\alpha(r-m), \qquad g(r)=k.
\label{drift_ou}
\end{equation}
The process is thus governed by Eq~(\ref{sde_1}). Letting $r_{0}=r(0)$ be the initial return, in the large time limit the probability density function 
$p(r,t|r_{0})$ has mean $m$ and variance $\sigma^{2}=k^{2}/2\alpha$.

\subsection{Adding risk aversion}

The Local Expectation Hypothesis (LEH) \cite{Cox1981,Gilles} assumes that prices can reasonably be modeled based on the data generating measure $p$. This results is attributed to the fact that inverstors are assumed to be risk neutral. However, if investors are sensitive to risk, bonds are priced with an artificial probability density function, $p^*$, called either the {\it risk-neutral probability measure} or the {\it risk-correcting measure} (see Appendix \ref{AppMPR}). The two measures $p$ and $p^*$ are related by the market price of risk (see below), which is the extra return per unit of risk that investors demand to bear risk. 

Following a standard procedure for bond pricing \cite{Vasicek,Duffie,Piazzesi,mmpdg_entropy_2022} we take risk into account by adjusting the drift term according to
\begin{equation}
f(r)= -\alpha(r-m) +g(r) q( r),
\label{f*}
\end{equation}
where in our case $g(r)=k>0$ and $q=q( r) \ge 0$ is the {\it market price of risk.} which, in general, may also depend on the current time, $q=q(r,t)$, but we here assume stationarity. The introduction of the market price of risk raises the effective interest rate, which in the context of bond pricing means that investors are compensated for taking increased risk. The most common assumption consists in taking $q(r,t)$ independent of $r$ and $t$, that is 
$$
q(r,t)=q= {\rm constant}.
$$
In such a case the adjusted drift can be rewritten as 
\begin{equation}
f( r)=-\alpha(r-m^*),
\label{f*_ou2}
\end{equation}
where
\begin{equation}
m^*=m + \frac{qk}{\alpha}.
\label{m*}
\end{equation}
The result is that the effective mean interest rate $m^*$ is increased relative to the historically observed interest rate $m$ by a constant amount that depends on the volatility, the reversion rate, and the market price of risk.

For the case of constant price of risk, the joint PDF is found by solving the Fokker-Planck equation (\ref{fpe_1}) , which now becomes (cf. Eqs (\ref{f*_ou2}))
\begin{equation}
\frac{\partial p}{\partial t}=-r\frac{\partial p}{\partial x}+\alpha\frac{\partial}{\partial r}\bigl[(r-m^*)p\bigr]+
\frac{1}{2}k^2\frac{\partial^{2} p}{\partial r^{2}},
\label{fpe_ou}
\end{equation}
with the initial condition provided in Eq. (\ref{initial_1}).
The equation problem can be addressed with the characteristic function (Fourier transform)
\begin{equation}
\tilde{p}(\omega_{1},\omega_{2},t|r_{0}) = \int_{-\infty}^{\infty}
e^{-i\omega_{1}x}dx \int_{-\infty}^{\infty}e^{-i\omega_{2}r}p(x,r,t|r_{0})dr.
\label{cf}
\end{equation}
Equation (\ref{fpe_ou}) can therefore take a simpler form:
\begin{equation}
\frac{\partial\tilde{p}}{\partial t}=(\omega_{1}-\alpha\omega_{2})\frac{\partial\tilde{p}}{\partial\omega_{2}}-
\left(im^*\alpha\omega_{2}+\frac 12 k^{2}\omega_{2}^{2}\right)  \tilde{p},
\label{cf_ou}
\end{equation}
with (cf. Eq (\ref{initial_1}))
\[
\tilde{p}(\omega_{1},\omega_{2},0|r_{0})=e^{-i\omega_{2}r_{0}}.
\]
The solution is given by
\begin{equation}
\tilde{p}(\omega_{1},\omega_{2},t) = 
\exp\Bigl\{-A(\omega_{1},t)\omega_{2}^{2}- B(\omega_{1},t)\omega_{2}-C(\omega_{1},t)\Bigr\},
\label{gaussian}
\end{equation}
where the expressions for the functions $A(\omega_{1},t)$, $B(\omega_{1},t)$, and
$C(\omega_{1},t)$ are obtained in Appendix \ref{AppA}. Comparing Eqs (\ref{D}) and (\ref{cf}) we obtain
\begin{equation}
D(t)=\tilde{p}\bigl(\omega_{1}=-i,\omega_{2}=0,t\bigr).
\label{discount_OU}
\end{equation}
Finally, it can be shown in Appendix \ref{AppA} that $D(t)=\exp\{-C(-i,t)\}$ which is a tractable expression for the discount rate (cf. Eq (\ref{a9})).   

We now make a change of notation. We have so far assumed that the discount functions is computed for a time $t$ in the future, starting at a fixed initial time $t_0 = 0$ with initial interest rate $r_0$ for a time $t$ in the future (let us remind that we have set $t_0=0$; had we kept $t_0\neq 0$ then time $t$ would have been replaced by $t-t_0$). However in what follows we will want to evaluate the discount starting from different initial times.  Recalling that the whole process is invariant under time translations, we can thus perform the change of notation such that $t_0 \rightarrow t$, $r_0 \rightarrow r(t)$ and $t-t_0 \rightarrow \tau$. Note that now $t$ is the present time, $r(t)=r$ is the present interest rate and the maturity time is $t+\tau$, that is to say, $\tau$ is the ``time to maturity''. The discount function is then (cf. Eq (\ref{a9}) of  Appendix \ref{AppA})
\begin{eqnarray}
\ln D(\tau)&=&-\frac{r}{\alpha}\left(1-e^{-\alpha \tau}\right)- 
m^* \left[\tau- \frac{1}{\alpha}\left(1-e^{-\alpha \tau}\right)\right]\nonumber \\
&+&\frac{k^{2}}{2\alpha^{3}}\biggl[\alpha \tau - 2\left(  1-e^{-\alpha \tau}\right)  +
\frac{1}{2}\left(1-e^{-2\alpha \tau}\right)\biggr].
\label{D_OU}
\end{eqnarray}
When the time to maturity $\tau$ is small we can approximate $D(\tau)$ by expanding the exponentials to first order using a Taylor series approximation and this yields 
$\ln D(\tau) \approx r\tau$ ($\tau\to 0$), as expected.

In the limit $\tau \rightarrow \infty$ we see that 
$$
\ln D(\tau)\approx [-m^*+k^2/(2\alpha^2)]\tau
$$
and the discount function becomes independent of the initial rate. It decays exponentially and can be written in the form
\begin{equation}
D(\tau)\simeq e^{-r_{\infty}\tau},
\label{assymptotic_D}
\end{equation}
where
\begin{equation}
r_\infty=m+\frac{qk}{\alpha}-\frac{k^2}{2\alpha^2},
\label{D_OU*_3(a)}
\end{equation}
is the long-run rate.

In general the long-run rate may be defined by 
\begin{equation}
r_\infty=-\lim_{\tau\to\infty}\frac{\ln D(\tau)}{\tau},
\label{r_infty_def}
\end{equation}
as long as the limit exists. Note that Eq (\ref{D_OU*_3(a)}) can be readily obtained after substituting Eq (\ref{D_OU}) into Eq (\ref{r_infty_def}).

From Eq (\ref{D_OU*_3(a)}) we see that the long-run discount rate depends on the historical rate $m$, but this is shifted by two terms.  The first shift raises the long-run rate due to the market price of risk. The second shift lowers it by an amount given by the ratio of uncertainty (as measured by $k$) and persistence (as measured by $\alpha$).   We can trivially rewrite the equation above as
\begin{equation}
r_\infty=m + \frac{k}{\alpha}\left(q - \frac{k}{2\alpha}\right).
\label{D_OU_2}
\end{equation}
This makes it clear that whether or not the overall shift in the long-run discount rate is positive or negative depends on the size of the market price of risk in relation to the ratio of the volatility parameter and the reversion rate.   

It is not surprising that the market price of risk raises the long term rate, but it is not so obvious that uncertainty and persistence can lower it.  Indeed for the OU process it can make it arbitrarily small.  For any given mean interest rate $m$, by varying $k$ and $\alpha$, the long-run discount rate $r_{\infty}$ can take on any value less than $m$, including negative values, while at the same time the
standard deviation $\sigma$ can also be made to take on any arbitrary positive value. In particular, by choosing the appropriate $(k,\alpha),$ we can make $r_{\infty}$ arbitrarily far below $m$ and $\sigma$ arbitrarily small. The
probability that $r(t)<r_\infty$ can be arbitrarily small, even when
$r_{\infty}\ll m$ (see   Appendix
 \ref{AppA}). Deducing the
correct parameters $(m,\sigma)$ of the stationary distribution of short run
interest rates does not determine $r_{\infty}$ by itself; on the contrary, any
$r_{\infty}<m$ is consistent with them. To infer $r_{\infty}${
from the data one must also tease out the mean reversion parameter }$\alpha.$
Holding the long run distribution $(m,\sigma)$ constant, by raising the
persistence parameter $1/\alpha$ it is possible to lower $r_{\infty}$ to any
desired level.  Of course, this can always be offset by the market price of risk.

It is even possible for the long-run rate to be negative.  This is due to the amplification of negative real interest rates $r(t)$.
Computation of the discount function involves an average over exponentials, rather than the exponential of an
average. As a result, periods where interest rates are negative are
amplified, and can easily dominate periods where interest rates are large and
positive, even if the negative rates are rarer and weaker. It does not take
many such periods to substantially reduce the long run interest rate.

To summarize, in the OU model the long-run discounting rate can be much lower than the mean,
and indeed can correspond to low interest rates that are rarely observed. 

\section{Empirical results}
\label{empirical_estimates}

\subsection{Estimation of real interest rates}

We estimate real interest rates as nominal rates corrected by inflation which we perform directly through the application of Fisher's equation by subtracting realized inflation from nominal interest rates (see Eq (\ref{real_rate_2}) below). This is not, however, the only way of obtaining real rates. Thus, for instance, Freeman et al. \cite{Freeman_2015}, among others, pursue an alternative procedure using cointegration methods to tease out real rates. Another way of obtaining real rates would be modeling nominal rates and inflation separately; that is, nominal rates by some positive random process (for example the Feller process as in the CIR model \cite{Farmer_etal_2015,Cox}) and the OU process for inflation since the latter can assume positive and negative values. Such a procedure --which unfortunately enlarge the number of parameters to be estimated from data-- would also need to take into account possible correlations between bond prices and inflation. It is not clear a priori which procedure is better, and the direct procedure we follow has the virtue of being much simpler with lesser number of free parameters to estimate.

We transform the annual rates into logarithmic rates and denote the resulting time series by $y(t|\tau)$ (with the maturity time $\tau$ equal to either 3 month or 10 years). Nominal rates $n(t)$ are then estimated by $n(t)\sim y(t|\tau)$ (see  Appendix  \ref{AppMPR} for details). The inflation rate $i(t)$ is estimated through the Consumer Price Index (CPI) as
\begin{equation}
i(t)\sim \frac 1\tau \ln\left[\frac{I(t+\tau)}{I(t)}\right] ,
\label{i(t)}
\end{equation}
where $I(t)$ is the aggregated inflation up to time $t$, and $\tau=10$ years (cf.  Appendix \ref{AppMPR}).  Finally, the real interest rate $r(t)$ is defined by Fisher's equation:

\begin{equation}
r(t)=n(t)-i(t).
\label{real_rate_2}
\end{equation}

We now estimate the OU model for real interest rates from historical data.  To this end we have collected long historical time series for both short and long run nominal interest rates, as well as inflation, for the United Kingdom and the United States.  The properties of the data are summarized in Table \ref{tab1}. For each country we have both three month and ten year interest rates, as well as an inflation index.

\begin{table}[!ht]
\begin{adjustwidth}{-2.25in}{0in} 
\begin{center}
\caption{\bf Data sets used in our main analysis for the United Kingdom (UK) and the United States (US).\label{tab1}}
\begin{tabular}{|l+l|l|l|l|r|}
\hline
{\bf Country} & {\bf Time series} & {\bf Frequency} &	{\bf From}	&	{\bf To}	& {\bf \# records} \\\thickhline
UK & 3 month Treasury bills  & monthly &12/31/1900	&12/31/2012	&113\\	\hline		
UK & 10 year bonds  & annual &12/31/1694	&12/31/2012	&309\\ \hline
UK & inflation index  & annual &	12/31/1694	&12/31/2012	& 309 \\ \hline
US & 3 month Treasury bills & monthly & 01/31/1920	&10/30/2012 &93 \\ \hline
US & 10 year bonds  & annual &12/31/1820	&10/30/2012	&183\\ \hline
US & inflation index  & annual & 12/31/1820	&10/30/2012	&183 \\ \hline
\end{tabular}
\end{center}
\end{adjustwidth}
\end{table}

\begin{figure}[!h]
\centering
\includegraphics[width=0.90\textwidth]{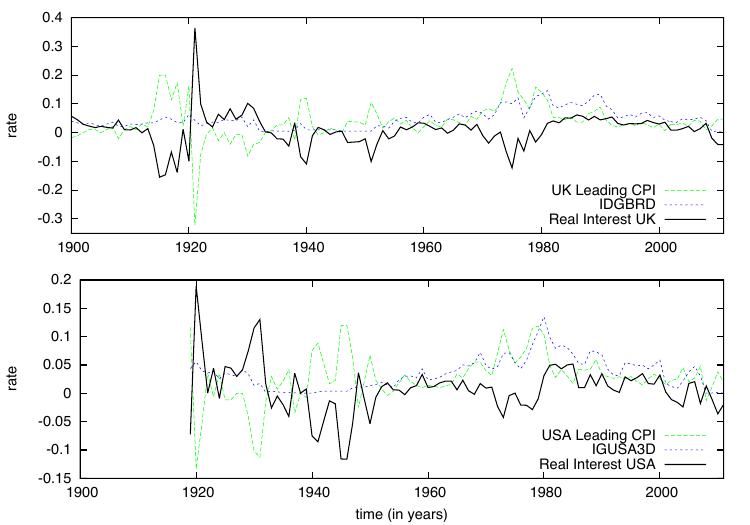}
\caption{{\bf (Color online)  Time series for the three month treasury bills for the UK (top) and US (bottom).} The inflation index is shown in green (long dashes), the nominal interest rate in blue (short dashes), and the real interest rate is shown as a black solid line. Real interest rates display large fluctuations and negative rates are
not uncommon.\label{fig1}}
\end{figure}

\begin{figure}[!h]
\centering
\includegraphics[width=0.9\textwidth]{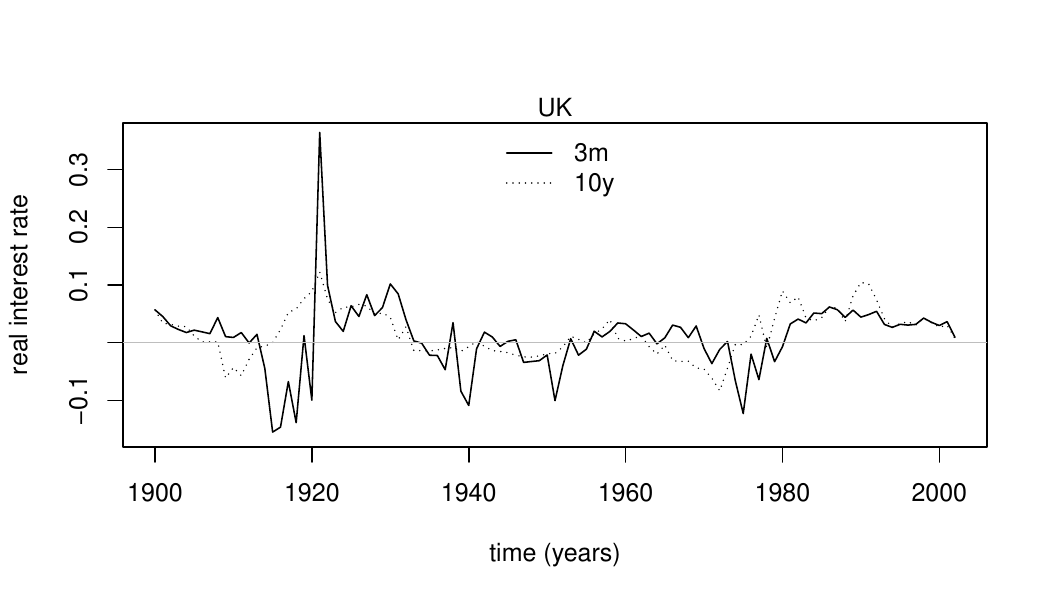}
\includegraphics[width=0.9\textwidth]{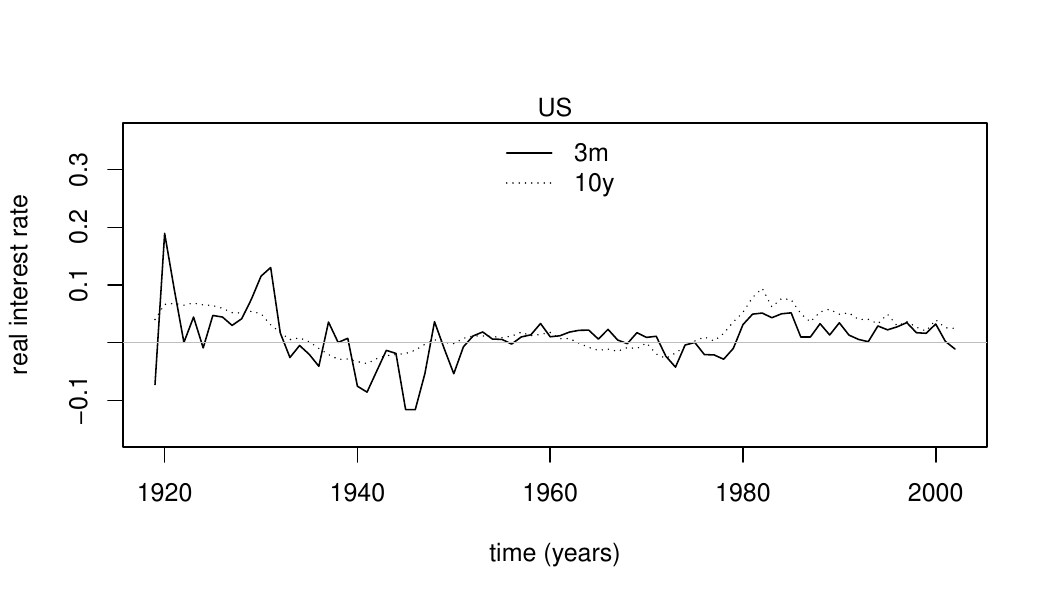}
\caption{{\bf Comparison between three month and ten year real interest rates for the UK (top) and the US (bottom).} The ten year real interest rates are shown with dashed lines and the three month real interest rates taken from Figure~\ref{fig1} are shown with solid lines.  A horizontal line is drawn at zero to make it clear when real rates are negative.\label{fig2}}
\end{figure}

The recording frequency for each country is either annual or monthly. For ten year government bonds (which pay out over a period of ten years) we smooth inflation rates with a ten-year forward moving average, and subtract the annualized inflation index from the annualized nominal rate to compute the real interest rate. For the three-month bond rates, in contrast, we use the inflation adjustment for the corresponding year (since we do not have inflation adjustments at quarterly frequency). Figure \ref{fig1} shows nominal rates, inflation and real interest rates for 3 month bonds for the UK and the US, and Figure \ref{fig2} compares 3 month and 10 year real interest rates. This procedure assumes that people correctly forecast inflation, thus, in absence of any knowledge of behavioral bias, we are assuming perfect rationality.

\subsection{Empirical properties of the data}

\begin{figure}
\begin{tabular}{cc}
\includegraphics[width= .45\textwidth]{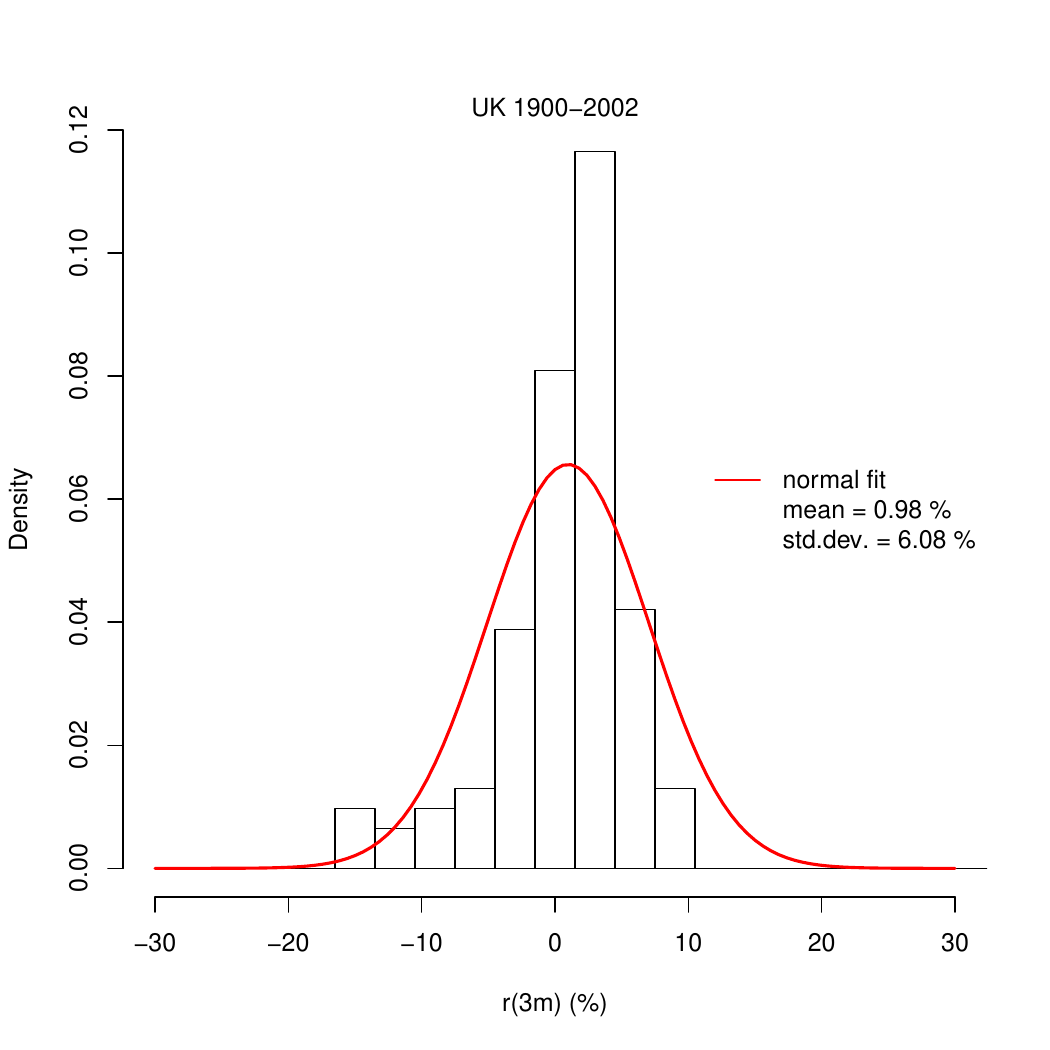} &
  \includegraphics[width=.45\textwidth]{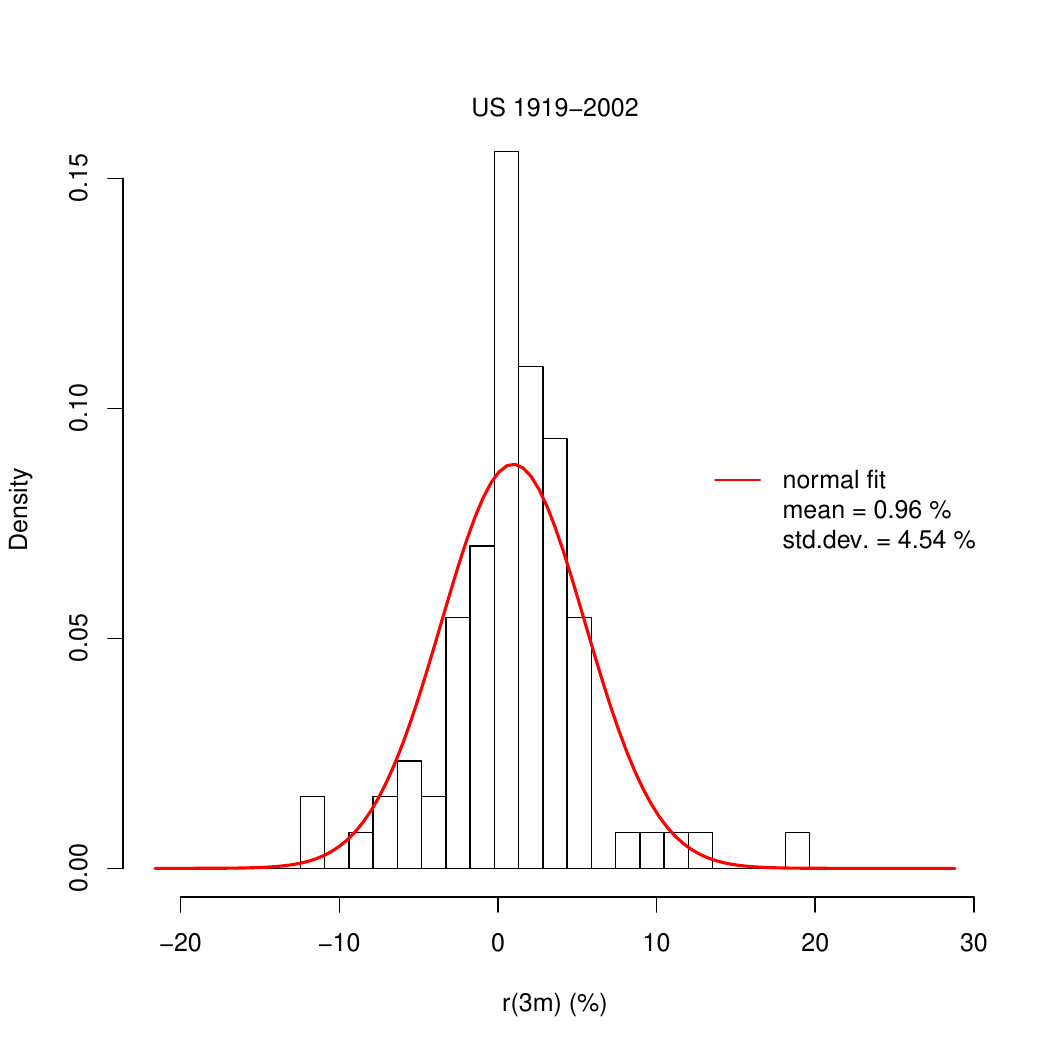} \\
  \includegraphics[width= .45\textwidth]{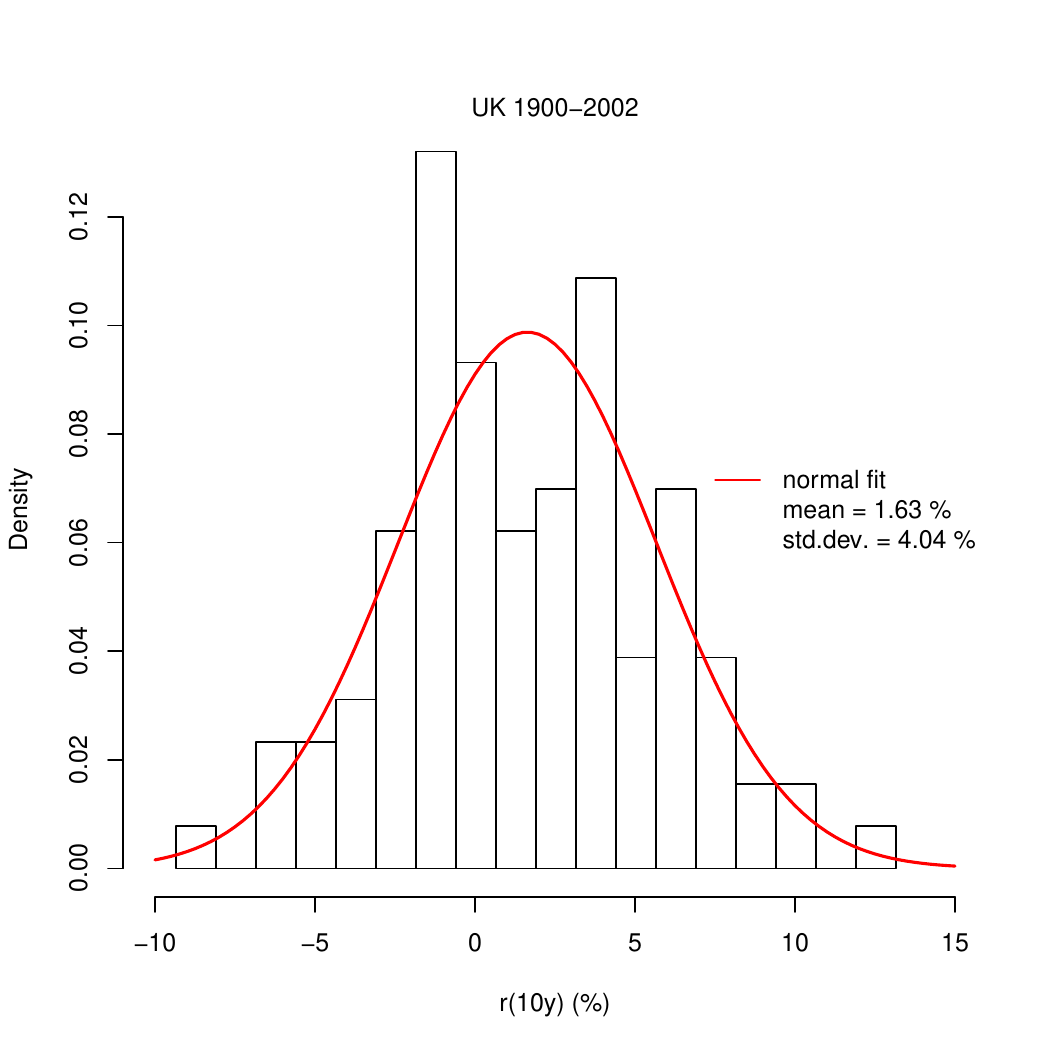} &
  \includegraphics[width= .45\textwidth]{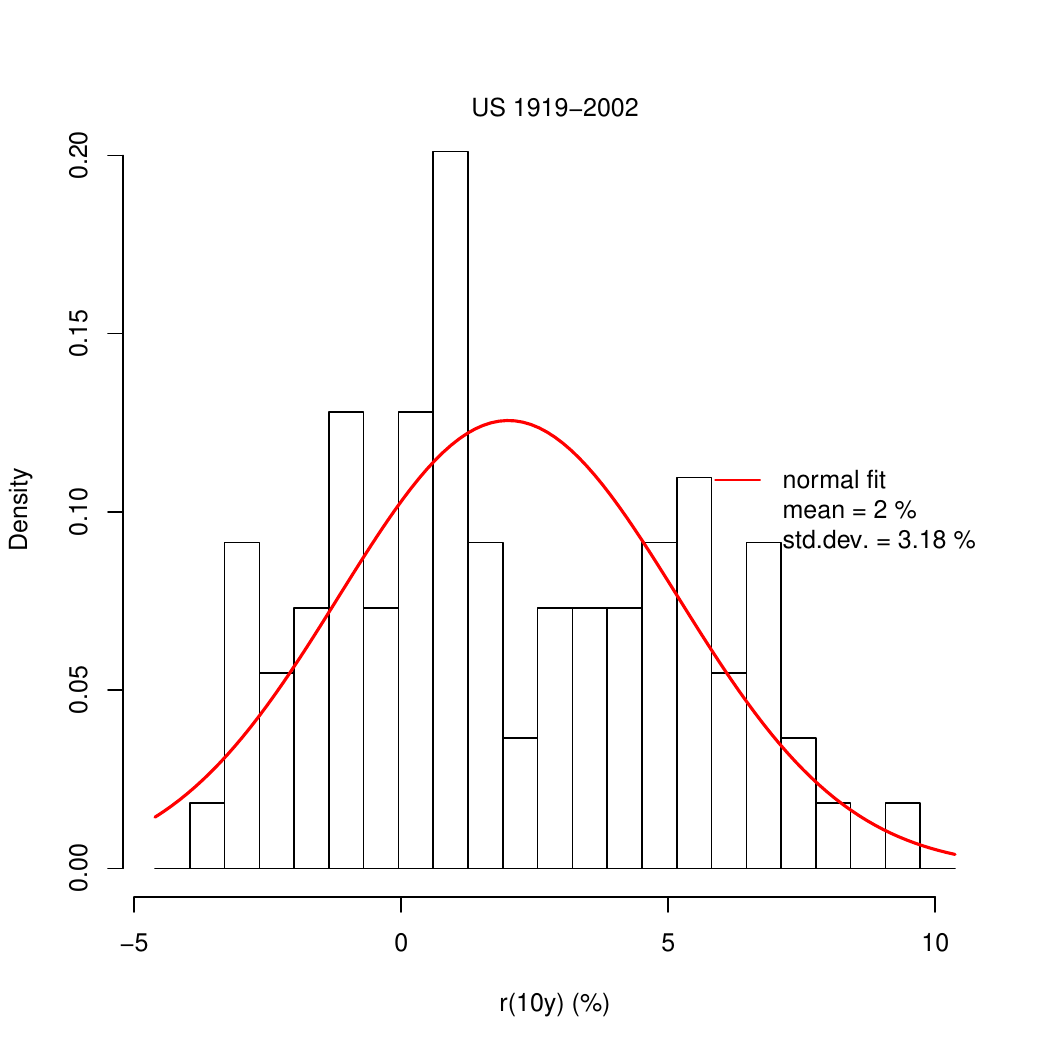}
\end{tabular}
\caption{{\bf Histograms of 3 month (top) and 10 year (bottom) real interest rates for the UK and the US.} The curve compares to a normal distribution.  This illustrates that negative real interest rates are common and that the distribution of interest rates is heavy-tailed relative to a normal distribution. \label{fig3}}
\end{figure}

One of the most striking features of these time series is that real interest rates are often negative, in some cases by substantial amounts and for long periods of time (see Figures \ref{fig1} and \ref{fig2}). This is evident looking at the histograms shown in Figure~\ref{fig3}.

Real 3 month interest rates for the UK are negative $32$ \% of the time, and there are four distinct periods where they drop nearly reach $-10$ \% and even below. 10 year real interest rates for the UK are negative $38$ \% of the time.  For the US real 3 month interest rates are also negative $32$ \% of the time, dropping below $-10$ \% during World War II, and 10 year real interest rates are negative $30$ \% of the time.  Given that real interest rates are negative about a third of the time, this makes it clear that models such as the log-normal process or the CIR model \cite{Farmer_etal_2015,Cox} that assume rates to be non-negative are far from being appropriate.  We therefore confine our empirical work to the Ornstein-Uhlenbeck model.

From Figure~\ref{fig3} it is also evident that the distribution of interest rates is heavy tailed.  This is particularly true for the short term interest rates.  Heavy tails are apparent because of the excess in the center of the distribution but also because the observations in the tail that exceed the normal distribution.  However these  deviations are not extreme and the OU process (which has a normal distribution) is at least a reasonable first approximation.   

Another striking feature is that the yield curve is intermittently inverted, that is, the 10 year real interest rate is at intervals lower than the 3 month rate (see  Figure \ref{fig1}). For the UK the yield curve is inverted slightly more than $50$\% of the time and for the US it is inverted $32$ \% of the time. Inversions of the yield curve are obviously important for understanding very long term rates.

\subsection{Parameter estimation}

The OU model with constant price of risk has four parameters to be estimated: $m$, $k$, $\alpha$ and $q$.   The 3 month rates are much less sensitive to the risk parameter $q$ than the 10 year rates. Therefore, by making the approximation that the 3 month is equivalent to the instantaneous process, we can estimate $m$, $k$ and $\alpha$ from the 3 month rate time series alone (see below for the estimation of $q$). The parameters obtained in this way are shown in Table \ref{tab2} and the values displayed are based on the maximum likelihood estimators derived in Ref. \cite{Brigo} (see Appendix \ref{AppC}).

To provide an appreciation for the robustness of the estimated parameters, we divide the time series into four blocks of equal size and evaluate the parameters $m$ and $k$ separately for each block, with the exception of the parameter $\alpha$ which is always estimated using the complete data set beacuse the time series in each block are too short for a reliable estimation of $\alpha$. The quoted uncertainties in $\alpha$ are then simply the standard least square error computed when fitting an exponential autocorrelation function of the real interest time series. The maximum and minimum values obtained for each parameter is listed in Table \ref{tab2}. The variations are large, indicating some combination of autocorrelation, non-stationarity and heavy tails.  

We remind that $m^*$ implicitly depends on the market price of risk $q$ (cf. Eq (\ref{m*})) and, as reported in Table \ref{tab2}, the value of $m^*$ for a very short time window ($\tau=3$ months) can be approximated by $m$. To estimate $q$, we thus make use of the $\tau=10$ year real interest rate series as it is considered to be a large value of $\tau$ (much larger than 3 months). The averages of the 3 month and 10 year interest when considering Eq~(\ref{D_OU}) give us two equations with two different values for $m^*$ where $k$ and $\alpha$ are hold fixed. And, finally, the market price of risk $q$ can be obtained with Eq (\ref{m*}). For this estimation we are also assuming in Eq~(\ref{D_OU}) that $r(t) \simeq m$, i.e., that the mean historical interest rate is equal (or very close) to the current rate.

\begin{table}[!ht]
\begin{center}
\caption{\bf Raw estimates of parameters of the risk-free Ornstein-Uhlenbeck model for the United Kingdom (UK) and the United States (US).\label{tab2}}
\begin{tabular}{|l+l|c|c+l|c|c+l|c|c|}
\hline
{\bf Country}	& {\bf $m^*$} & {\bf Min} & {\bf Max} & {\bf $k$} & {\bf Min} & {\bf Max} & {\bf $\alpha$} & {\bf Min}  & {\bf Max} \\ \thickhline
UK & $0.88$ & $-0.4$ & $3.5$ & $8.2$ & $1.8$ & $15.6$ & $0.93$ & $0.2$ & $1.4$ \\ \hline
US & $0.83$ & $-1.2$ & $2.2$ & $5.7$ & $2.5$ & $10.5$ & $0.74$ & 0.3 & $1.3$  \\
\hline
\end{tabular}
\begin{flushleft} Results are based on the annually sampled time series of three-month real interest rates.We use annual units. The Min and Max columns correspond to the minimum and the maximum values of the parameters obtained by splitting the time series into four blocks of equal length and estimating the parameters separately in each block (except $\alpha$, see main text). For better estimates see Table \ref{tab3}.
\end{flushleft}
\end{center}
\end{table}

The estimates obtained in this manner are slightly distorted not only because the 3 month bond is sampled annually but also because we have treated it as though it were an instantaneous rate.  We can estimate the size of this bias by simulating the instantaneous process, which we approximate as having daily frequency.  The simulation procedure is standard, we thus generate the instantaneous process $r(t)$ by solving numerically the OU equation with parameters $\alpha$, $m$, and $k$ given by the empirical values taken from 3 month rates as given in Table \ref{tab2} (see more details in  Appendix \ref{AppC}). We next create a surrogate time series for the 3 month real interest rate time series using Eq~(\ref{D_OU}) with $\tau = 0.25$ year and the empirical $r=r(t)$ as the initial condition at each time $t$.  We then mimic the procedure employed for the real data by estimating the parameters based on the surrogate 3 month rate, sampled at annual frequency.  We can thus adjust the parameters of the instantaneous process to roughly match, on average, those observed for the real data (so that the estimated values based on the surrogate 3 month series match those of the observed 3 month series).  The parameters with the bias corrected are given in Table \ref{tab3} (see Appendix \ref{AppC} for more details on the procedure we followed in order to correct the bias).  The resulting shift in parameters is small, as can be seen by comparing Tables \ref{tab2} and \ref{tab3}. The main difference is in the parameter $\alpha$, which sets the timescale for mean reversion; this changes by a little more than $10$ \%. Our numerical experiments indicate that the main source of the bias is the annual sampling and since parameter $\alpha$ sets the timescale for mean reversion it is not surprising that it is affected by this.  

\begin{table}[!ht]
\begin{adjustwidth}{-2.25in}{0in}
\begin{center}
\caption{\bf Parameters of the instantaneous Ornstein-Uhlenbeck process.\label{tab3}}
\begin{tabular}{|l+l|c|c+l|c|c+lc|c+l|c|c|}
\hline
{\bf Country}	& {\bf $m$} & {\bf $5$\% }& {\bf $95$\%} & {\bf $k$} & {\bf $5$\%} & {\bf $95$\%} & {\bf $\alpha$} & {\bf $5$\%} & {\bf $95$\%} & {\bf $q$} & {\bf $5$\%} & {\bf $95$\%} \\ \thickhline
UK & $0.84$ & $-0.92$ & $2.6$ & $8.9$ & $7.6$ & $10.4$ & $0.82$ & $0.47$ & $1.26$ & $0.13$ & $-0.04$  & $0.34$ \\ \hline
US & $0.83$ & $-0.85$ & $2.3$ & $5.8$ & $4.9$ & $6.8$ & $0.65$ & 0.36 & $1.06$ & $0.20$ & $0.02$ & $0.43$ \\
\hline
\end{tabular}
\begin{flushleft} We use the procedure described in the text. Parameters of the instantaneous Ornstein-Uhlenbeck process using the procedure described in the text. $m$ and $k$ are in percent.
\end{flushleft}
\end{center}
\end{adjustwidth}
\end{table}

\begin{table}[!ht]
\begin{adjustwidth}{-2.25in}{0in}
\begin{center}
\caption{\bf A comparison of the percentage of the time real interest rates are negative for both the UK and the US and for both the data and the model simulation. \label{tab4}}
\begin{tabular}{|l+c|c|c|c|}
\hline
{\bf Country} 	& {\bf 3 month (data)} & {\bf 3 month (model)} & {\bf 10 year (data)} & {\bf 10 year (model)} \\ \thickhline
UK & $32$ \% & $43$ \% & $38$ \% & $34$ \%\\ \hline
USA	& $32$ \% & $42$ \% & $30$ \% & $26$ \% \\
\hline
\end{tabular}
\begin{flushleft} The simulation values are averaged over 1\,000 simulations.
\end{flushleft}
\end{center}
\end{adjustwidth}
\end{table}

\begin{figure}
\centering
\includegraphics[width=0.9\textwidth]{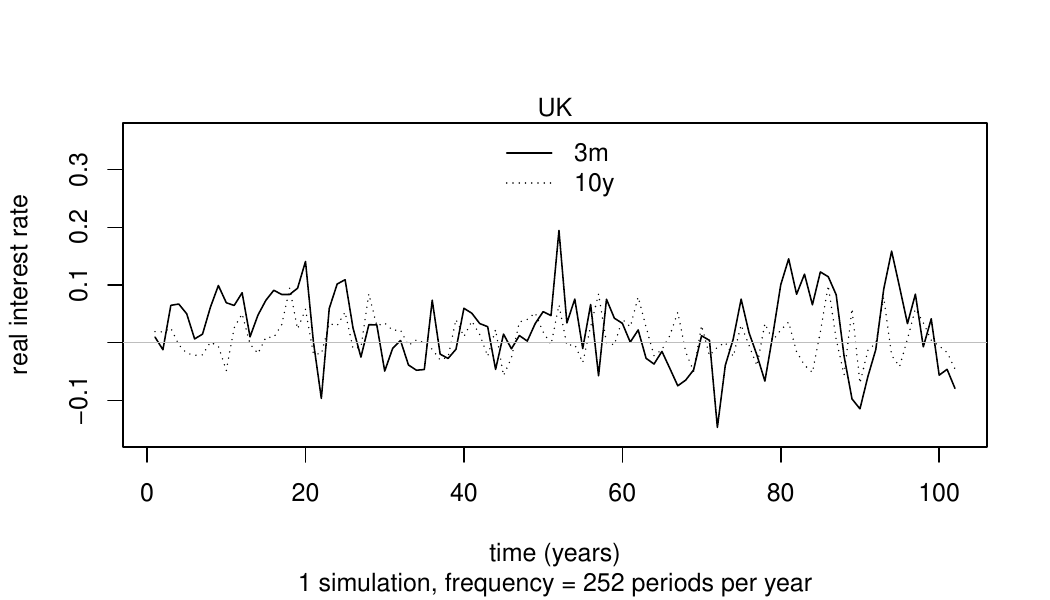}
\includegraphics[width=0.9\textwidth]{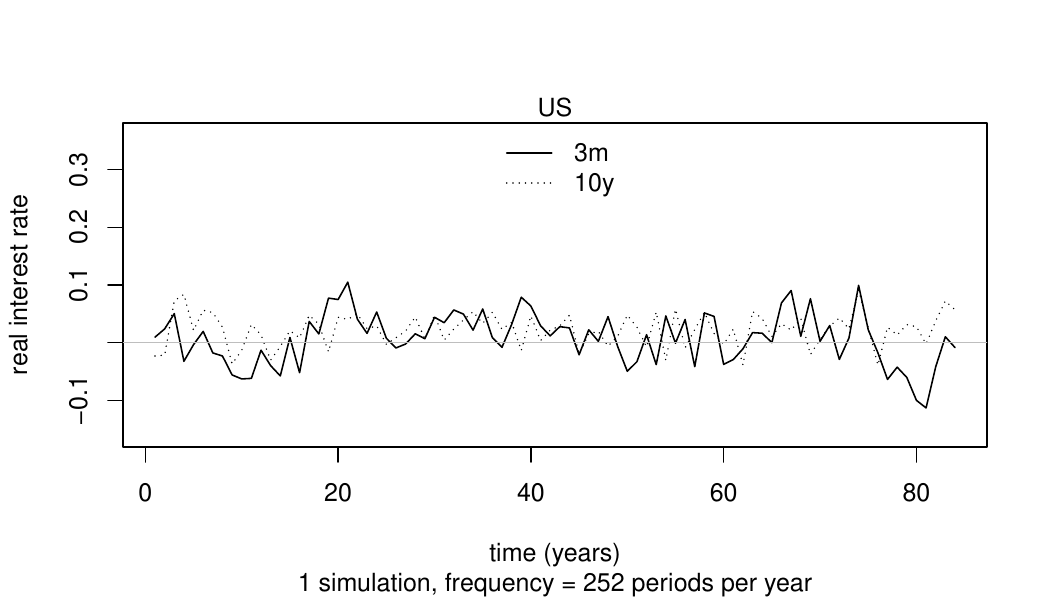}
\caption{{\bf A simulation of the 3 month and 10 year interest rates for the UK (above) and the US (below) using the OU process.} Compare with Figure \ref{fig2}. \label{fig4}}
\end{figure}

\subsection{Contrasting the model with data}

We now make some comparisons of the simulated OU model with real data (see Table \ref{tab4}). Similarly to the procedure for the 3 month rates, we create simulated 10-year rates using Eq~(\ref{D_OU}), so that rates are defined as $(1/\tau)\ln D(\tau)$ with $\tau= 10$ year, where $r=r(t)$ as the initial condition at each time $t$ (recall that maturity time is $t+\tau$). A comparison shows that the standard deviation of the 3 month rates matches reasonably well, but the standard deviation of the simulated rates is much lower than that of the 10 year rates.   We correct this by adding IID normally distributed noise to match the standard deviation of the simulated series with real data.  We also neglect the 10 year smoothing of the 10 year inflation data.  The simulated result has a lower correlation between 3 month and 10 year rates than real data; for the UK the correlations are $21$ \% (simulated) vs. $39$ \% real and for the US $24$ \% (simulated) vs. $59$ \% (real).  However, the distributions for both 3 month and 10 years agree reasonably well. Figure \ref{fig4} shows the simulated 3 month and 10 year interest rate time series, which should be compared to the real data shown in Figure \ref{fig2}. Not surprisingly, since the 3 month simulated rates are normally distributed, they lack the extreme values observed in the real data.

The OU model does a good job of capturing the frequency of negative interest rates and yield curve inversions. Table \ref{tab4} compares the frequency of negative interest rates for the real data and the simulation for both 3 month and 10 year rates.

\begin{figure}
\centering
\includegraphics[width=1\textwidth]{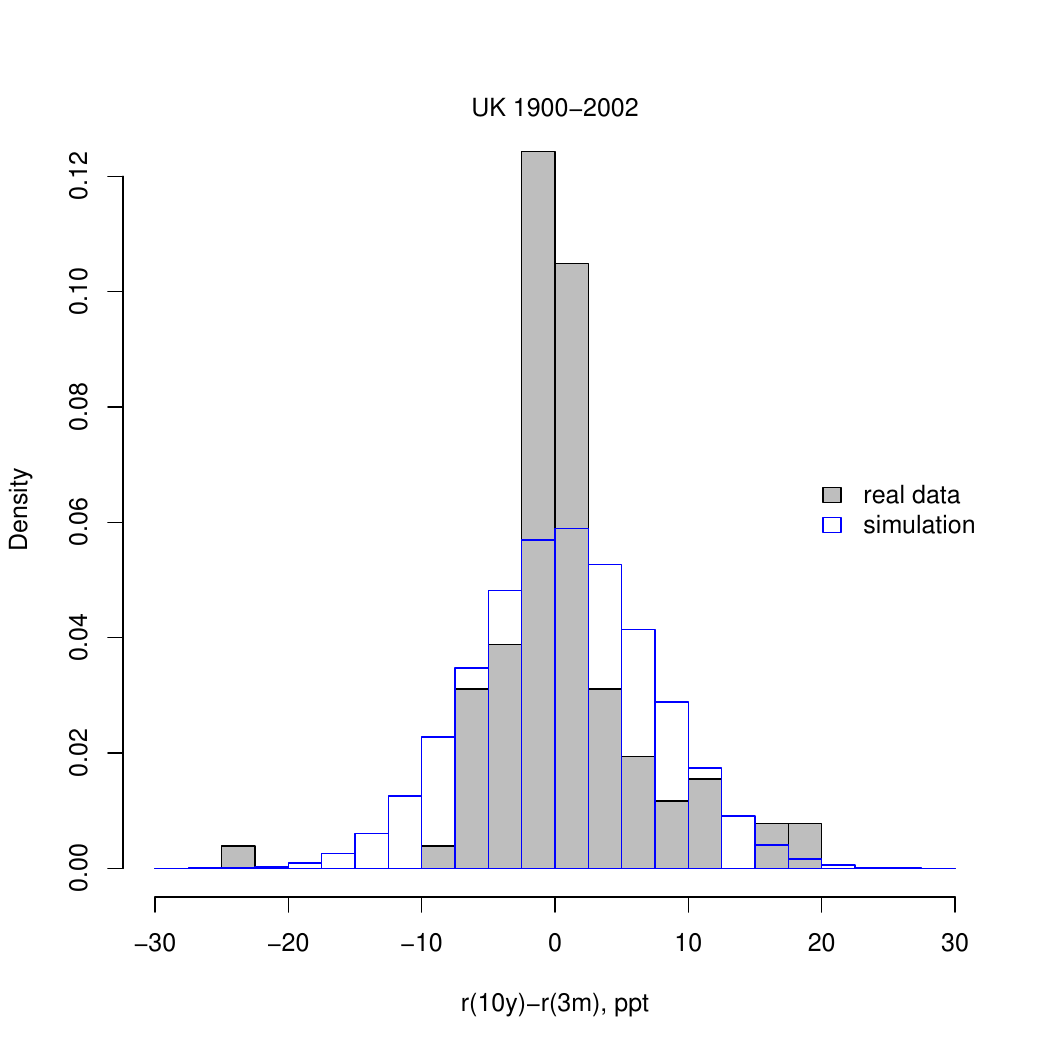}
\caption{{\bf (Color online)  Histogram comparing yield curve inversions in the simulated vs. real data for the UK.} We measure yield curve inversion based on the difference between the 10 year interest rate and the three month interest rate; positive values indicate a normal yield curve and negative values an inverted yield curve.  Interest rates are measured in percent.   The real data are shown in grey, the simulation in white.  The real data are heavier tailed, but the agreement is otherwise reasonably good.\label{fig5}}
\end{figure}

In Figure~\ref{fig5} we present a histogram of yield curve inversions for both the data and the model for the UK. The US histogram is qualitatively similar and we do not present it here. We use the difference between the 10 year real interest rate and the 3 month interest rate as our measure of inversion.  The inversions of the data are somewhat more heavy tailed than those of the model, but the agreement is surprisingly good. The real UK yield curve is inverted roughly $50$ \% of the time and the simulated yield curves are inverted $46$ \% of the time. Similarly the real US yield curves is inverted $32$ \% of the time and the simulated yield curves are inverted $41$ \% of the time.

\subsection{Estimating confidence intervals}

With such a short time series as we have, it is difficult to estimate confidence intervals by methods such as bootstrapping. This is particularly true for $\alpha$, where the time series properties of the data matter, so that one would need to do a block bootstrap and unfortunately there are not many blocks of sufficient length. However, assuming that the model is well-specified we can at least compute error estimates which are consistent with our model. The width of the resulting confidence intervals can be regarded as lower bounds on the width of the confidence intervals, and provide a perception for the magnitude of the estimating errors.  

We repeatedly simulate the instantaneous process $r(t)$ using the parameters estimated from the data and generate 3 month and 10 year series as described above. In order to accurately mimic the constraints imposed by the data we sample the simulated series at an annual frequency. We then apply the estimation procedure described above to estimate the four parameters. Doing this 1\,000 times allows us to compute the $5$ \% and $95$ \% quantiles for each parameter. The results are shown in Table \ref{tab3}.
  
\begin{figure}
\centering
\includegraphics[width=0.45\textwidth]{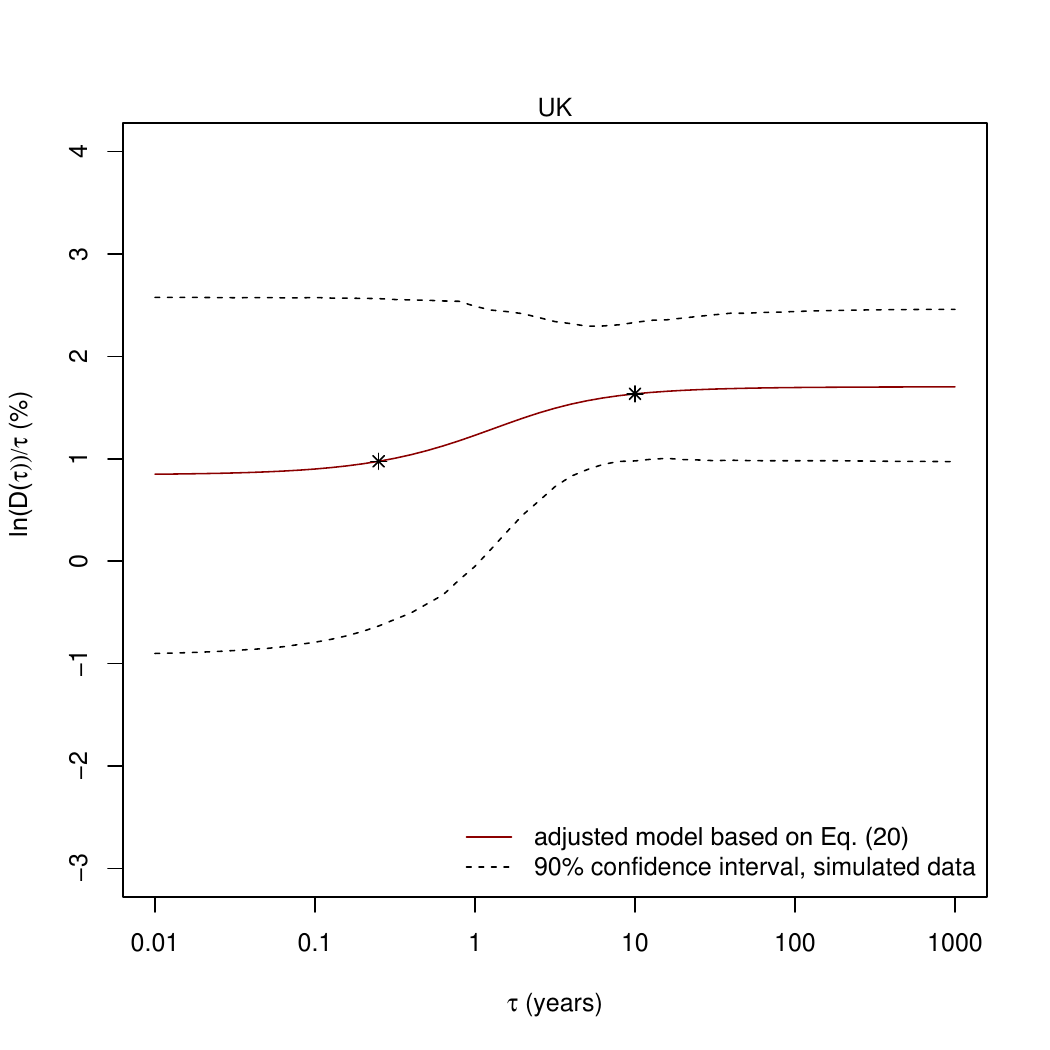}
\includegraphics[width=0.45\textwidth]{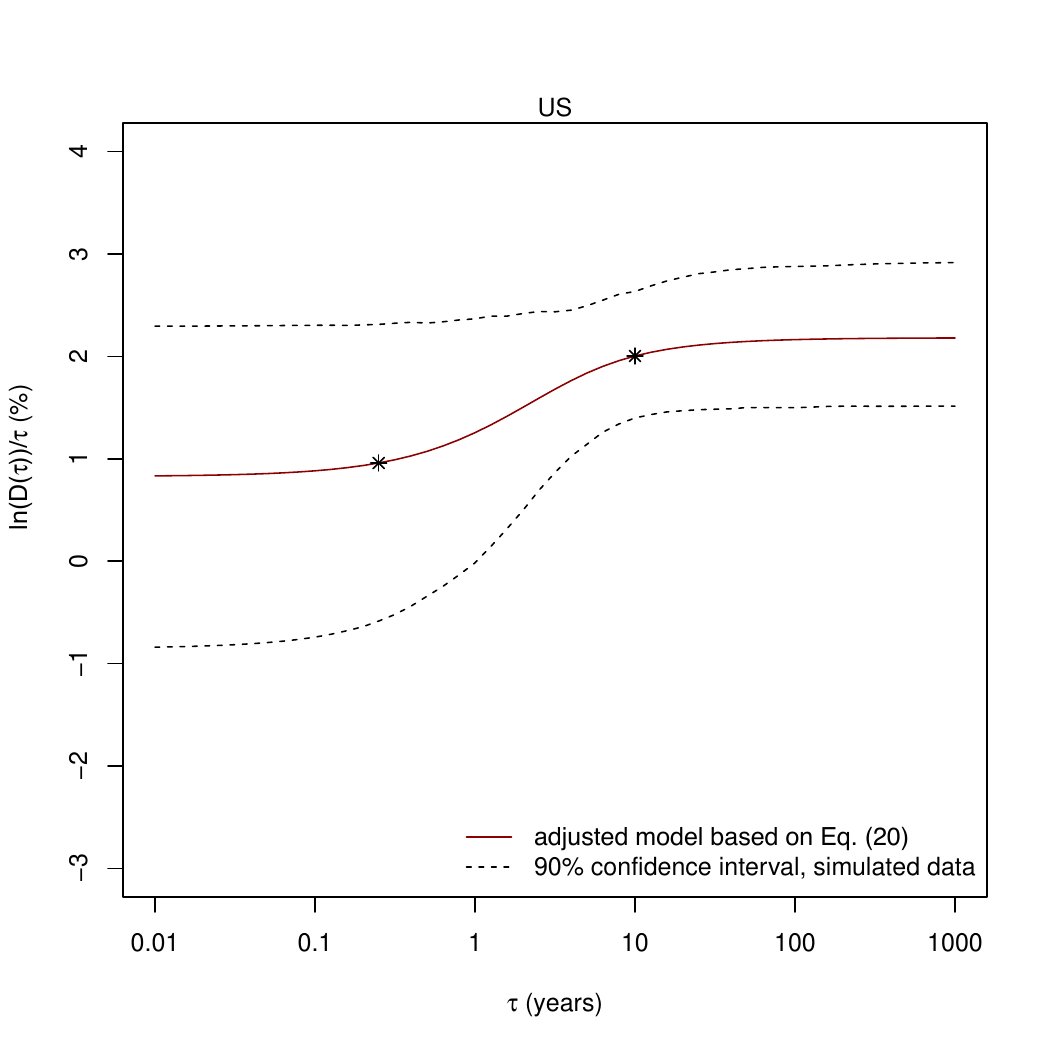}
\caption{{\bf (Color online) The estimated discount rate $\ln(D(\tau))/\tau$ (in \%) for the UK (top) and US (bottom)}. The rates are shown as solid red lines (based on Eq (20)), plotted versus time to maturity $\tau$ (time is on logarithmic scale). The dashed lines indicate the $5$ \% and $95$ \% quantiles based on the simulation procedure described in the text. These values are obtained with 1,000 simulated series, each 84 years long, with 252 periods per year. Starred points are the empirical estimated values from UK and US historical time series.\label{fig6}}
\end{figure}

The estimated discount functions, together with their confidence intervals, are shown in Figure~\ref{fig6}. The uncertainty intervals are estimated by repeatedly simulating the instantaneous, 3 month and 10 year processes as described above, applying the estimation procedure to the simulated data, and computing the discount function at each time interval. This is repeated 1\,000 times to estimate the $5$ \% and $95$ \% quantiles.

We are finally ready to present our key result. The long term interest rate $r_\infty$ is computed using  Eq~(\ref{D_OU*_3(a)}) based on the values in Table \ref{tab3} and the final key result of this paper is presented in Table \ref{tab5}. The mean long run rate is $r_\infty = 1.69$ \% for the UK and $2.21$ \% for the US. Let us note that because of the scarcity of the data available uncertainties are substantial, with standard deviations of roughly $0.45$ \% in both cases.

\begin{table}[!ht]
\begin{center}
\caption{\bf Long term interest rate $r_\infty$ for the United Kingdom (UK) and the United States (US).}
\begin{tabular}{|l+r|r|r|}
\hline
{\bf Country} 	& {\bf $r_\infty$} &{\bf $5$\%} & {\bf $95$\% }\\ \thickhline
UK & $1.69$ & $0.76$ & $2.63$ \\ \hline
USA	& $2.21$ & $1.35$ & $3.07$\\
\hline
\end{tabular}
\begin{flushleft} Measured in percent, as well as the $5$ \% and $95$ \% quantiles.
\end{flushleft}
\label{tab5}
\end{center}
\end{table}

\section{Discussion}

Climate change and climate action are widely studied from a variety of perspectives \cite{Soergel2021}. However, researchers in financial economics have only recently turned their attention to climate change and climate finance is currently a quickly growing research field \cite{Giglio_2021}. Finance academics participating in a recent survey have identified discount as a key topic to reduce climate risks \cite{Stroebel_2021}. With this paper, we have taken a muldisciplinary perspective from complex systems science and their related methods to study historical bond prices \cite{Bauer2021} with the aim to contribute to the need of new approaches to evaluate the climate action urgency \cite{Stern2022,Stern2022b}.

More specifically, we have wanted to infer that long term discount rates is not just a trivial matter of extrapolating mean interest rates, but rather one must take several non-trivial factors into account. To begin with, because real interest rates are so often substantially negative, one should use a model that permits negative rates. This leads us to the Ornstein-Uhlenbeck model.  While the presence of negative rates in this model may be viewed as a liability for describing nominal rates, for real rates this becomes a virtue.  Another factor that should be considered is the market price of risk, which tends to raise longer term rates.  Finally one should properly take into account the uncertainty and persistence of interest rates, which tends to lower the long-term discount rate.  The use of the OU model accommodates all of these factors.  When we estimate the OU model and compare it with real data, we see a good agreement on several essential properties, such as the frequencies of negative rates and yield curve inversions.

Our results indicate that the long term interest rate used by Stern \cite{Stern} is supported by historical data. His value of $1.4$ \% is less than a standard deviation below the estimated long term rate for the UK of $1.69$ \%, and just under two standard deviations of the US long term rate of $2.21$ \%. More recent estimates by Stern indeed pose an scenario with a value below 1 \% \cite{Stern2022}. In contrast, higher long run rates --see, for instance, \cite{Nordhaus2007,Nordhaus}-- seem not to be supported, as they are well above the $95$ \% confidence intervals of $2.63$ for the UK and $3.07$ for the US. Our estimates of $1.69$ \% (UK) and $2.2 $ \% (US) are compatible with the rates recently estimated by Giglio et al. (\cite{giglio_2015,giglio_2016,giglio_2021}) which use data from UK housing markets during 2004--2013  and Singapore during 1995--2013 to estimate an annual discount rate of 2.6 \% for payments more than 100 years in the future.

Our results could potentially be improved on in several ways. One would be to acquire more data.  This could include data for more countries --as we did recently but without considering market price of risk in a recent publication \cite{perello_etal_JSM}-- as well as longer term bonds or inflation-indexed bonds. Another possible improvement would be to extend the model to better capture the nonstationarity and/or heavy tailed behavior observed in the data \cite{Gollier2008}. In this regard we suspect that had we were able to take the observed nonstationarity \cite{Stern2022} and/or heavy tails into account, mean values would have decreased because of the boosting of the uncertainty/persistence effect. This possibility is under present investigation.

In a broader level, both Stern and Stiglitz have recently provided new methods and models which could be further studied from a complex systems science perspective \cite{Stern2022b}. Stern \cite{Stern2022} has indeed put the accent in incorporating stronger multidisciplinary perspective to consider social and behavioral dimensions \cite{Li2022,Creutzig2020,Kundzewicz2020}. The same author \cite{Stern} also mentions a definition of social discounts that might include inputs from social dilemmas and predefined behavioral experiments \cite{Vicens2018}.

\section*{Conclusion}

To conclude, we have demonstrated that historical data indicates that the long-term discount rate is probably not very large.  While the error bars remain large, a value of $2$ \% or less seems plausible, corresponding to a present value of about $14$ \% for a payment received 100 years in the future. Other higher values do not seem to be fully consistent with the historical data. The need for immediate and substantial spending to combat climate change is thus sustained.

\section*{Appendices}

\appendix

\section{Discount function for the Ornstein-Uhlenbeck model}
\label{AppA}

We have seen in the main text that when rates are described by the OU process the joint characteristic function $\tilde p(\omega_1,\omega_2,t|r_0)$ obeys the first-order partial differential equation (cf Eq (\ref{cf_ou}))
\begin{equation}
\frac{\partial\tilde{p}}{\partial t}=(\omega_{1}-\alpha\omega_{2}
)\frac{\partial\tilde{p}}{\partial\omega_{2}}-\left( im\omega_{2}+\frac 12 k^{2}\omega_{2}^{2}\right) \tilde{p},
\label{a1}
\end{equation}
with initial condition ($t_0=0$)
\begin{equation}
\tilde{p}(\omega_{1},\omega_{2},0|r_{0})=e^{-i\omega_{2}r_{0}}.
\label{a2}
\end{equation}

Due to the linearity of the OU process and the Gaussian character of the input noise, we may look for a solution of the initial value problem (\ref{a1})-- (\ref{a2}) in the form of a Gaussian density:
\begin{equation}
\tilde p(\omega_1,\omega_2,t)=\exp\Bigl\{-A(\omega_1,t)\omega_2^2 - B(\omega_1,t)\omega_2-C(\omega_1,t)\Bigr\},
\label{a3}
\end{equation}
where $A(\omega_1,t)$, $B(\omega_1,t)$, and $C(\omega_1,t)$ are unknown functions to be consistently determined. Substituting Eq (\ref{a3}) into Eq (\ref{a1}), identifying like powers in $\omega_2$ and taking into account Eq (\ref{a2}), we find that these functions satisfy the following set of differential equations
\begin{equation}
\dot{A}=-2\alpha A-k^2/2, \qquad\qquad A(\omega_1,0)=0;
\label{a4}
\end{equation}
\begin{equation}
\dot B=-\alpha B+2\omega_1 A-im\alpha, \qquad B(\omega_1,0)=ir_0;
\label{a5}
\end{equation}
\begin{equation}
\dot C=\omega_1 B, \qquad\qquad\qquad\qquad C(\omega_1,0)=0.
\label{a6}
\end{equation}

Equation (\ref{a4}) is a first-order linear differential equation that can be readily solved giving 
\begin{equation}
A(\omega_1,t)=\frac{k^2}{4\alpha}\left(1-e^{-2\alpha t}\right),
\label{a7}
\end{equation}
substituting this expression for $A(\omega_1,t)$ into Eq (\ref{a5}) results in another first-order equation for $B(\omega_1,t)$, whose solution reads
\begin{equation}
B(\omega_1,t)=ir_0e^{-\alpha t}+\frac{k^2\omega_1}{2\alpha^2}\Bigl(1-2e^{-\alpha t}+e^{-2\alpha t}\Bigr)
+im\Bigl(1-e^{-\alpha t}\Bigr).
\label{a8}
\end{equation}
Finally, the direct integration of Eq (\ref{a6}) yields the expression for $C(\omega_1,t)$ 
\begin{eqnarray}
C(\omega_1,t)=i\omega_1r_0\frac{1}{\alpha}\left(1-e^{-\alpha t}\right) &+&\frac{k^2\omega_1^2}{2\alpha^3} 
\left[\alpha t-2\left(1-e^{-\alpha t}\right)+\frac 12\left(1-e^{-2\alpha t}\right)\right] \nonumber \\
&+&im\omega_1\left[t-\frac{1}{\alpha}\left(1-e^{-\alpha t}\right)\right].
\label{C}
\end{eqnarray}

From Eq (\ref{discount_OU}) we see that the effective discount is given by the characteristic function, $\tilde p(\omega_1,\omega_2,t|r_0)$, evaluated at the points $\omega_1=-i$ and $\omega_2=0$. Thus from Eqs (\ref{a3}) and (\ref{C}) we obtain
\begin{eqnarray}
\ln D(t) =-\frac{r_{0}}{\alpha}\left( 1-e^{-\alpha t}\right)&+&\frac{k^{2}}{2\alpha^3}\biggl[\alpha t-2\left(1-e^{-\alpha t}\right) 
+\frac{1}{2}\left( 1-e^{-2\alpha t}\right) \biggr] \nonumber \\
&-& m\left[ t-\frac{1}{\alpha}\left( 1-e^{-\alpha t}\right) \right] ,
\label{a9}
\end{eqnarray}
which, with the change of notation regarding time as explained in the main text, agrees with Eq (\ref{D_OU}).

Knowing the exact form of the probability distribution of the joint process $(x(t),r(t))$ --through its joint characteristic function (\ref{a3})-- it is possible study in detail some interesting properties of the rate $r(t)$, as we do next.

\subsection{Negative rates}

As pointed out in the main text, a characteristic of the OU model is the possibility of attaining negative values. This probability is given by 
\begin{equation}
P(r<0,t|r_0)=\int_{-\infty}^0 p(r,t|r_0)dr,
\label{a11}
\end{equation}
where $p(r,t|r_0)$ is the probability density function of the rate process. This is given by the marginal density
$$
p(r,t|r_0)=\int_{-\infty}^\infty p(x,r,t|r_0)dx,
$$
and the characteristic function of the rate is related to the characteristic function of the bidimensional process $(x(t),r(t))$ by the simple relation
$$
\tilde p(\omega_2,t|r_0)=\tilde p(\omega_1=0,\omega_2,t|r_0).
$$

From Eq (\ref{a3}) and Eqs (\ref{a7})-(\ref{C}) we get
$$
\tilde p(\omega_2,t|r_0)=\exp\Biggl\{-\frac{k^2}{4\alpha}\left(1-e^{-2\alpha t}\right)\omega_2^2
-i\left[r_0e^{-\alpha t}+m\left(1-e^{-\alpha t}\right)\right]\omega_2\Biggr\},
$$
which after Fourier inversion gives the Gaussian density
\begin{equation}
p(r,t|r_0)=\frac{(\alpha/k^2)^{1/2}}{\sqrt{\pi(1-e^{-2\alpha t})}}
\exp\left\{-\frac{(\alpha/k^2)[r-r_0e^{-\alpha t}-m(1-e^{-\alpha t})]^2}{1-e^{-2\alpha t}}\right\}. 
\label{pdf_r}
\end{equation}

The probability for $r(t)$ to be negative, Eq (\ref{a11}) then reads
\begin{equation}
P(r<0,t|r_0)=\frac 12{\rm Erfc}\left(\frac{(\alpha/k^2)^{1/2}[r_0e^{-\alpha t}+m(1-e^{-\alpha t})]}{\sqrt{1-e^{-2\alpha t}}}\right),
\label{p-1}
\end{equation}
where ${\rm Erfc}(z)$ is the complementary error function \cite{mos},
$$
{\rm Erfc}(z)=\frac{2}{\sqrt\pi} \int_z^\infty e^{-x^2}dx.
$$

Note that as time increases (in fact starting from $t>\alpha^{-1}$) the probability (\ref{p-1}) is well approximated by the stationary probability, defined as
$$
 P_s^{(-)}\equiv\lim_{t\rightarrow\infty}P(r<0,t|r_0).
$$ 
That is
\begin{equation}
P_s^{(-)}=\frac 12{\rm Erfc}\left(m\sqrt{\alpha/k^2}\right).
\label{p-stat}
\end{equation}

In terms of the dimensionless normal level $\mu$ and the dimensionless volatility $\kappa$ defined by 
\begin{equation}
\mu\equiv m/\alpha, \qquad \kappa\equiv k/\alpha^{3/2},
\label{mu}
\end{equation}
this probability reduces to
\begin{equation}
P_s^{(-)}=\frac 12{\rm Erfc}\left(\mu/\kappa\right).
\label{p-stat2}
\end{equation}
Let us now see the behavior of $P_s^{(-)}$ for the cases (i) $\mu<\kappa$ and (ii) $\mu>\kappa$. 

(i) If the normal rate $\mu$ is smaller than rate's volatility $\kappa$, we use the series expansion \cite{mos}
$$
{\rm Erfc}(z)=1-\frac{2}{\sqrt\pi}z+O(z^2).
$$
Hence,
\begin{equation}
P_s^{(-)}=\frac 12-\frac{1}{\sqrt\pi}(\mu/\kappa)+O(\mu^2/\kappa^2).
\label{p-i}
\end{equation}
For $\mu/\kappa$ sufficiently small, this probability approaches $1/2$. In other words, rates are positive or negative with almost equal probability. Note that this corresponds to the situation in which noise dominates over the mean. In the original units (cf. Eq (\ref{mu})) this case corresponds to the noise intensity $k$ being larger than $m\alpha^{1/2}$.

(ii) When fluctuations around the normal level are smaller than the normal level itself, $\kappa<\mu$, we use the asymptotic approximation \cite{mos}
$$
{\rm Erfc}(z)\sim\frac{e^{-z^2}}{\sqrt\pi z}\left[1+O\left(\frac{1}{z^2}\right)\right],
$$
and
\begin{equation}
P_s^{(-)}\sim\frac{1}{2\sqrt\pi}\left(\frac{\kappa}{\mu}\right)e^{-\mu^2/\kappa^2}.
\label{p-ii}
\end{equation}
Therefore, for mild fluctuations around the mean (that is, when $k\ll m\alpha^{1/2}$) the probability of negative rates is exponentially small.

\subsection{Rates below the long-run rate} 
 
It is also interesting to know the  probability that real rates $r(t)$ are below the long-run rate $r_\infty$. This is given by 
 $$
 P_\infty(t)\equiv {\rm Prob}\{ r(t)<r_\infty\}=\int_{-\infty}^{r_\infty}p(r,t|r_0)dr.
 $$
 In the stationary regime, $t\rightarrow\infty$, we have
\begin{equation}
P_\infty=\int_{-\infty}^{r_\infty}p(r)dr,
\label{P_inf}
\end{equation}
where $p(r)$ is the stationary PDF. For the OU model $p(r)$ is obtained from Eq (\ref{pdf_r}) after taking the limit $t\rightarrow\infty$:
\begin{equation}
p(r)=\frac{1}{\sqrt{\pi}}\left(\frac{\alpha}{k^2}\right)^{1/2} e^{-\alpha(r-m)^2/k^2}.
\label{p(r)}
\end{equation}
Substituting Eq (\ref{p(r)}) into Eq (\ref{P_inf}) we write
$$
P_\infty=\frac{1}{\sqrt{\pi}}\left(\frac{\alpha}{k^2}\right)^{1/2}\int_{-\infty}^{r_\infty} e^{-\alpha(r-m)^2/k^2}dr=
\frac{1}{\sqrt{\pi}}\left(\frac{\alpha}{k^2}\right)^{1/2}\int_{-r_\infty}^{\infty}e^{-\alpha(r+m)^2/k^2}dr,
$$
or in terms of the complementary error function
\begin{equation}
P_\infty=\frac 12 {\rm Erfc}\left[\frac{\sqrt\alpha}{k}(m-r_\infty)\right].
\label{P2}
\end{equation}
Using the asymptotic estimates of the complementary error function discusse above, we see that this probability is exponentially small if 
$(m-r_\infty) \to \infty$ with $\sqrt\alpha/k$ fixed, or if $\sqrt\alpha/k\to\infty$ with a fixed differential of rates $(m-r_\infty)$.

\section{Real and nominal rates. The market price of risk}
\label{AppMPR}

Recall that real rates are defined as the difference between nominal rates and inflation rates (cf. Eq (\ref{real_rate_2})):
$$
r(t)=n(t)-i(t).
$$
We now discuss how to estimate $n(t)$ and $i(t)$ from empirical data .

\subsection{Nominal rates}

Let $B(t|t+\tau)$ be the price at time $t$ of a government bond maturing at time $t+\tau$ ($\tau\geq 0$) with unit maturity, 
$B(t|t)=1$. The instantaneous rate of return, $b(t|t+\tau)$, of this bond is defined as 
$$
b(t|t+\tau)\equiv\frac{1}{B(t|t+\tau)}\frac{d B(t|t+\tau)}{dt},
$$
so that, 
\begin{equation}
B(t|t+\tau)=\exp\left[-\int_t^{t+\tau} b(t|t')dt'\right].
\label{n_1}
\end{equation}
It is also useful to define the ``yield to maturity'' $y(t|\tau)$ as 
$$
y(t|\tau)\equiv-\frac 1\tau \ln B(t|t+\tau),
$$
or, after using Eq (\ref{n_1}), as 
\begin{equation}
y(t|\tau)=\frac 1\tau \int_t^{t+\tau} b(t|t')dt'.
\label{n_2}
\end{equation}
This form of defining $y(t|\tau)$ has and interesting interpretation since it shows that the yield to maturity is the time average over the maturing period $\tau$ of the instantaneous rate of return.

Let us remark that the data at our disposal are not the historical values of $B(t|t+\tau)$ but the annual interest rate of the zero-coupon bond $\beta(t|\tau)$. In this case we have
\begin{equation}
B(t|t+\tau)=\frac{1}{[1+\beta(t|\tau)]^\tau},
\label{n_3}
\end{equation}
so that
\begin{equation}
y(t|\tau)=\ln[1+\beta(t|\tau)].
\label{n_4}
\end{equation}

The spot or nominal rate $n(t)$ is defined as
\begin{equation}
n(t)\equiv\lim_{\tau\to 0} y(t|\tau),
\label{n_5}
\end{equation}
which, after substituting for Eq (\ref{n_2}) yields
\begin{equation}
n(t)=b(t|t).
\label{n_5b}
\end{equation} 

When dealing with empirical data, nominal rates are thus estimated by the yield,  
\begin{equation}
n(t)\sim y(t|\tau)=\ln[1+\beta(t|\tau)]
\label{n_6}
\end{equation}
and, attending to definition (\ref{n_5}), the shorter $\tau$ is,  the better the estimation for $n(t)$.

\subsection{Inflation rates}

$I(t)$ is defined as the aggregated inflation up to time $t$. The inflation rate $i(t)$ can be then estimated by the {\it ex post} mean inflation rate over a period of time $\tau$, $i(t|\tau)$:
\begin{equation}
i(t|\tau)\equiv\frac 1\tau \ln\frac{I(t+\tau)}{I(t)},
\label{i_1}
\end{equation}
Also, the Consumer Price Index (CPI, $C(t)$) is related to the aggregated inflation 
\begin{equation}
I(t+\tau)=I(t)\prod_{j=0}^{\tau-1}\bigl[1+C(t+j)\bigr].
\label{i_2}
\end{equation}
Finally, the instantaneous rate of inflation $i(t)$ can be estimated by $i(t+\tau)$ and in terms of the CPI
\begin{equation}
i(t)\sim i(t+\tau)=\frac 1\tau \sum_{j=0}^{\tau-1} \ln\bigl[1+C(t+j)\bigr]
\label{i_3}
\end{equation}

\subsection{The market price of risk}

The concepts of risk neutral probabilities and market price of risk (MPR) were developed for bonds and nominal rates. They can be, nonetheless, extended formally to real rates despite practical difficulties which arise because real rates are not tradable and, thus, an empirical basis for constructing a risk neutral measure is lacking.     

Let us recall that real rates $r(t)$ are estimated by the quantity $r(t|\tau)$:
\begin{equation}
r(t)\sim r(t|\tau)\equiv y(t|\tau)-i(t|\tau),
\label{r_1}
\end{equation}
where $y(t|\tau)$ is the yield to maturity $\tau$ for a zero-coupon bond $B(t|t+\tau)$ and $i(t|\tau)$ is the inflation rate over period $\tau$. 

From theoretical point of view the instantaneous real rate $r(t)$ is defined by 
$$
r(t)=\lim_{\tau\to 0}r(t|\tau).
$$
This leads us to take the shortest possible yield, $y(t|\tau)$, at our disposal ($\tau=3$ months) to construct a proxy of the real spot rate $r$.  

Obviously the spot rate $r(t)$ is random, so is the quantity $r(t|\tau)$. We denote by $\mu$ and $\sigma^2$ the average and variance of $r(t|\tau)$ respectively. From empirical data these statistics are estimated by
$$
\mu\sim\frac 1N\sum_{t=1}^Nr(t|\tau), \qquad \sigma^2\sim\frac 1N\sum_{t=1}^N\bigl[r(t|\tau)-\mu\bigr]^2,
$$
where $N$ is the number of samples.
Note that in the most general situation $\mu=\mu(t,r|\tau)$ and  $\sigma=\sigma(t,r|\tau)$ depend on current time $t$, rate $r$, and maturing interval $\tau$ \cite{Vasicek}. 

The risk premium is defined by the difference $\mu(t,r|\tau)-r$. Since this excess return depends on the maturity time there can be arbitrage opportunities by buying and selling bonds at different maturities \cite{Vasicek} (see also \cite{maso_llibre}). It can be shown that these arbitrage opportunities are ruled out as long as the Sharpe ratio of the excess return,
\begin{equation}
q(r,t)\equiv\frac{\mu(t,r|\tau)-r}{\sigma(t,r|\tau)},
\label{mpr_def}
\end{equation}
is independent of the maturity time $\tau$ \cite{Vasicek}. This ratio is called the market price of risk. 
It depends in general of the current time $t$ and the spot rate $r$ although the most common and feasible assumption is that $q$ is constant or, at most, a function of $r$ (see main text). 

\section{Parameter estimation and uncertainties}
\label{AppC}

\subsection{Parameter estimation}

Let us recall that the OU model is defined by means of the linear stochastic differential equation:
$$
dr(t)=-\alpha(r-m)dt+kdW(t)
$$
whose solution is 
$$
r(t)=r(t_0)e^{-\alpha(t-t_0)}+m\left[1-e^{-\alpha(t-t_0)}\right] +k\int_{t_0}^te^{-\alpha(t-t')}dW(t'),
$$
where $t_0$ is an arbitrary initial time. In what follows we will assume that the process is in the stationary regime. That is to say, we assume the process started at the infinite past (i.e., $t_0=-\infty$) and at present time $t$ all transient effects have faded away. Therefore,
\begin{equation}
r(t)=m+k\int_{-\infty}^te^{-\alpha(t-t')}dW(t').
\label{R}
\end{equation}
The parameter $m$ is easily estimated by noting that since the Wiener process has zero mean (stationary) rate
\begin{equation}
{\mathbb E}[r(t)]=m.
\label{m}
\end{equation}

The parameters $\alpha$ and $k$ can be estimated in terms of the correlation function
$$
K(t-t')= {\mathbb E}\left[(r(t)-m)(r(t')-m)\right].
$$
From Eqs (\ref{R}) and (\ref{m}) we write
$$
K(t-t')=k^2e^{-\alpha(t+t')}\int_{-\infty}^t e^{\alpha t_1}
\int_{-\infty}^t e^{\alpha t_2}{\mathbb E}[dW(t_1)dW(t_2)].
$$
Taking into account that \cite{maso_llibre}
$$
{\mathbb E}[dW(t_1)dW(t_2)]=\delta(t_1-t_2)dt_1dt_2,
$$
where $\delta(\cdot)$ is the Dirac delta function, and performing the integration on $t_2$, we have
$$
K(t-t')=k^2e^{-\alpha(t+t')}\int_{-\infty}^t \Theta(t'-t_1)e^{2\alpha t_1} dt_1,
$$
where $\Theta(\cdot)$ is the Heaviside step function. In the evaluation of the integral we must take into account whether $t>t'$ or $t<t'$. It is a simple matter to see that for either case the final result is
\begin{equation}
K(t-t')=\frac{k^2}{2\alpha} e^{-\alpha|t-t'|}.
\label{corr}
\end{equation}

Let us incidentally note that Eq (\ref{corr})  proves that the correlation time of the OU process is given by $\alpha^{-1}$. Indeed, the correlation time, $\tau_c$, of any stationary random process with correlation function $K(\tau)$ is defined by the time integral of $K(\tau)/K(0)$. In our case
\begin{equation}
\tau_c\equiv\frac{1}{K(0)}\int_0^\infty K(\tau)d\tau=\frac 1\alpha.
\label{corr_time}
\end{equation}
The empirical auto-correlation fit to an exponential (cf. Eq (\ref{corr})) allows to estimate $\alpha$. 

The parameter $k$ can finally be obtained from the (empirical) standard deviation
$$
\sigma^2={\mathbb E}\left[(r(t)-m)^2\right],
$$
Hence,
\begin{equation}
k=\sigma\sqrt{2\alpha}.
\label{k}
\end{equation}

We estimate these quantities for the three month interest rates using the maximum likelihood procedure given in \cite{Brigo}.

\subsection{Correcting for the bias using 3 month rates sampled at annual frequency}

Our parameter estimation process that corrects for the bias introduced by using the 3 month rate as an approximation to the instantaneous rate, and by sampling it at annual frequency, has the following steps:
\begin{enumerate}
\item
Estimate parameters using the historical 3 month and 10 year data as described in the main text.
\item
Simulate the instantaneous process (which we approximate as a daily process) using the parameters inferred in step (1) to generate a simulated time series $r(t)$ whose length matches that of the real data (roughly 100 years for the UK and 80 years for the US).  
\item
Construct simulated 3 month and 10 year time series based on Eq~(\ref{D_OU}) with $\tau = 0.25$ and $\tau = 10$, using the time series for $r(t)$ from step 1 as the initial condition for each time $t$.
\item
Estimate $m$, $k$ and $\alpha$ on the simulated 3 month series (sampled at annual frequency).  
\item
Repeat steps (2-4) for 1000 times and compute the average value of each parameter under the estimation process of step (4).  This yields systematic shifts in the parameters relative to those estimated on the historical data, making it clear that the estimation process is biased. 
\item
Correct for this bias by adjusting the parameters of the instantaneous process by the magnitude of the average shift, so that the estimation process for the simulated 3 month bond times series roughly matches the values estimated from the historical series.   
\end{enumerate}

\section*{Acknowledgments}
This work was partially supported by Ministerio de Ciencia e Innovación (MCIN, Spain), Agencia Estatal de Investigación (AEI) AEI/10.13039/501100011033 and Fondo Europeo de Desarrollo Regional (FEDER) [grant number PID2019-106811GB-C33, MM, JP and JM]; by Generalitat de Catalunya (Spain) through Complexity Lab Barcelona [grant numbers 2017 SGR 608 and 2021 SGR 00856; MM, JP and JM].

\nolinenumbers

%
%
%

\end{document}